\documentclass[showpacs,preprintnumbers,amsmath,amssymb]{revtex4}

\usepackage{epsfig}
\usepackage{rotating}
\usepackage{graphicx}
\usepackage{dcolumn}
\usepackage{bm}

\begin{document}

\title{
Nonlinear parametric model for Granger causality of time series}
\author{Daniele Marinazzo$^{1,2,3}$, Mario Pellicoro$^{1,2,3}$  and Sebastiano Stramaglia$^{1,2,3}$}

 \affiliation{
$^1$TIRES-Center of
Innovative Technologies for Signal Detection and Processing,\\
Universit\`a di Bari, Italy\\
$^2$Dipartimento Interateneo di Fisica, Bari, Italy \\
$^3$Istituto Nazionale di Fisica Nucleare, Sezione di Bari, Italy  }

\date{\today}

\begin{abstract}
We generalize a previously proposed approach for nonlinear Granger causality of time
series, based on radial basis function. The proposed model is not constrained to be
additive in variables from the two time series and can approximate any function of these
variables, still being suitable to evaluate causality. Usefulness of this measure of
causality is shown in a physiological example and in the study of the feed-back loop in a
model of excitatory and inhibitory neurons. 
\end{abstract}

\maketitle
\section{introduction}
Since the seminal paper by Granger \cite{granger}, detecting causality relationships
between two simultaneously recorded signals is one of the most important problems in time
series analysis. Applications arise in many fields, like economy \cite{ting}, brain
studies \cite{tass,arnold,sowa,rosemblum,blinowska,pereda}, human cardiorespiratory
system \cite{cardioresp}, and many others. The major approach to this problem examines if
the prediction of one series could be improved by incorporating information of the other,
as proposed by Granger. In particular, if the prediction error of the first time series
is reduced by including measurements from the second time series in the regression model,
then the second time series is said to have a causal influence on the first time series.
As Granger causality was originally developed for linear systems \cite{granger}, recently
some attempts to extend this concept to the nonlinear case have been proposed. In
\cite{chen} local linear models in reduced neighborhoods are considered and the average
causality index, over the whole data-set, is proposed as a nonlinear measure. In
\cite{ancona} a  radial basis function (RBF) approach has been used to model data, while
in \cite{verdes} a non-parametric test of causality has been proposed, based on the
concept of {\it transfer entropy}. A recent paper \cite{nc} pointed out that not all
nonlinear prediction schemes are suitable to evaluate causality between two time series,
since they should be invariant if statistically independent variables are added to  the
set of input variables. This property guarantees that, at least asymptotically, one would
be able to recognize variables without causality relationship. The purpose of this work
is to use the theoretical results found in \cite{nc} to find the largest class of RBF
models suitable to evaluate causality, thus  extending  the results described in
\cite{ancona}. Moreover we show the application of causality to the analysis of
cardiocirculatory interaction and to study the mutual influences in inhibitory and
excitatory model neurons.

Let $\{\bar{x}_i\}_{i=1,.,N}$ and $\{\bar{y}_i\}_{i=1,.,N}$ be two time series of $N$
simultaneously measured quantities. In the following we will assume that time series are
stationary. We aim at quantifying {\it how much} $\bar{y}$ {\it is cause of}  $\bar{x}$.
For $k=1$ to $M$ (where $M=N-m$, $m$ being the order of the model), we denote
$x^k=\bar{x}_{k+m}$, ${\bf X}^k=(\bar{x}_{k+m-1}, \bar{x}_{k+m-2},...,\bar{x}_{k})$,
${\bf Y}^k=(\bar{y}_{k+m-1}, \bar{y}_{k+m-2},...,\bar{y}_{k})$ and we treat these
quantities as $M$ realizations of the stochastic variables ($x$, ${\bf X}$, ${\bf Y}$)
\cite{nota1}. Let us now consider the general nonlinear model
\begin{eqnarray}
\begin{array}{l}
x=w_0 + {\bf w_{1}}\cdot {\bf \Phi}\left({\bf X}\right)+{\bf w_{2}}\cdot {\bf \Psi}\left({\bf Y}\right)+{\bf w_{3}}\cdot {\bf \Xi}\left({\bf X},{\bf Y}\right),\\
\label{mod-non}
\end{array}
\end{eqnarray}
where $w_0$ is the bias term, $\{\bf w\}$ are real vectors of free parameters, ${\bf
\Phi}=\left(\varphi_1,...,\varphi_{n_x}\right)$ are $n_{x}$ given nonlinear real
functions of $m$ variables, ${\bf \Psi}=\left(\psi_1,...,\psi_{n_y}\right)$ are $n_y$
other real functions of $m$ variables, and ${\bf
\Xi}=\left(\xi_1,...,\xi_{n_{xy}}\right)$ are $n_{xy}$ functions of $2m$ variables.
Parameters $w_0$ and $\{\bf w\}$ must be fixed to minimize the prediction error (we
assume $M \gg 1+n_{x}+n_{y}+n_{xy}$):
\begin{eqnarray}
\begin{array}{l}
\epsilon_{xy}={1\over M}\sum_{k=1}^M \left(x^k-w_0-{\bf w_{1}}\cdot {\bf \Phi}\left({\bf X}^k\right)-{\bf w_{2}}\cdot {\bf \Psi}\left({\bf Y}^k\right)+{\bf w_{3}}\cdot {\bf \Xi}\left({\bf X}^k,{\bf Y}^k\right)\right)^2.\\
\end{array}
\end{eqnarray}
We also consider the model:
\begin{eqnarray}
\begin{array}{l}
x=v_0+{\bf v_{1}}\cdot {\bf \Phi}\left({\bf X}\right),\\
\end{array}
\label{mmod}
\end{eqnarray}
and the corresponding prediction error $\epsilon_{x}$. If the
prediction of $\bar{x}$ improves by incorporating the past values of
$\{\bar{y}_i\}$, i.e. $\epsilon_{xy}$ is smaller than
$\epsilon_{x}$, then $y$ is said to have a causal influence on $x$.
We must require that, if ${\bf Y}$ is statistically independent of
$x$ and ${\bf X}$, then $\epsilon_{xy}=\epsilon_{x}$ at least for
$M\to\infty$. This is ensured if, for each $\alpha\in
\{1,\ldots,n_{xy}\}$,  an index $i_\alpha$ exists such that:
\begin{equation}
\xi_\alpha \left({\bf X},{\bf Y}\right)=\varphi_{i_\alpha}({\bf X})\Gamma_\alpha({\bf
Y}), \label{cond} \end{equation} where $\Gamma_\alpha$ is an arbitrary function of
$\mathbf{Y}$. As explained in \cite{nc}, model (\ref{mod-non}), with condition
(\ref{cond}), is the largest class of nonlinear parametric models suitable to evaluate
causality. We remark that $\epsilon_{xy}$ is equal to $\epsilon_{x}$ also at finite $M$,
for statistically independent $\mathbf{Y}$, if the probability distribution is replaced
by the empirical measure. Exchanging the two time series, one may analogously study the
causal influence of $x$ on $y$.

We choose the functions ${\bf \Phi}$, ${\bf \Psi}$ and ${\bf \Xi}$, in model
(\ref{mod-non}), in the frame of RBF methods thus generalizing the approach in
\cite{ancona}. We fix $n_x =n_y =n_{xy}=n\ll M$: $n$ centers $\{{\bf \tilde{X}}^\rho,{\bf
\tilde{Y}}^\rho\}_{\rho=1}^n$, in the space of $({\bf X},{\bf Y})$ vectors, are
determined by a clustering procedure applied to data $\{({\bf X}^k,{\bf Y}^k)\}_{k=1}^M$
(we use fuzzy c-means \cite{fcm} to find prototypes). We then make the following choice
for $\rho=1,\ldots,n$:
\begin{eqnarray}
\begin{array}{ll}
\varphi_\rho \left({\bf X}\right)&=\exp\left({-\|{\bf X}-{\bf \tilde{X}}^\rho\|^2/2\sigma^2}\right),\\
\psi_\rho \left({\bf Y}\right)&=\exp\left({-\|{\bf Y}-{\bf \tilde{Y}}^\rho\|^2 /2\sigma^2}\right),\\
\xi_\rho \left({\bf X},{\bf Y}\right)&=\varphi_\rho \left({\bf X}\right)\psi_\rho
\left({\bf Y}\right),
\end{array}
\label{eq-rbf}
\end{eqnarray}
$\sigma$ being a fixed parameter, whose order of magnitude is the
average spacing between the centers. Condition (\ref{cond}) is
satisfied by construction. The RBF model of \cite{ancona} is
recovered setting $\bf{\Xi} =0$ in (\ref{mod-non}), i.e. it is
constrained to be additive in variables $\textbf{X}$ and
$\textbf{Y}$; instead the RBF model, here proposed, can approximate
any function of $\textbf{X}$ and $\textbf{Y}$.

\begin{figure}[ht!]
\begin{center}
\epsfig{file=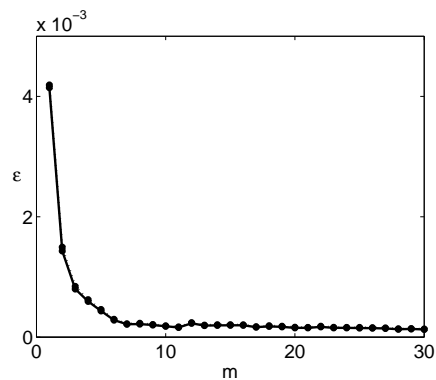,height=5.cm}
\end{center}
\caption{{\small $\epsilon_x$ and $\epsilon_y$ (indistinguishable) are plotted vs $m$,
for a typical patient.\label{figepsvsm}}}
\end{figure}

\begin{figure}[ht!]
\begin{center}
\epsfig{file=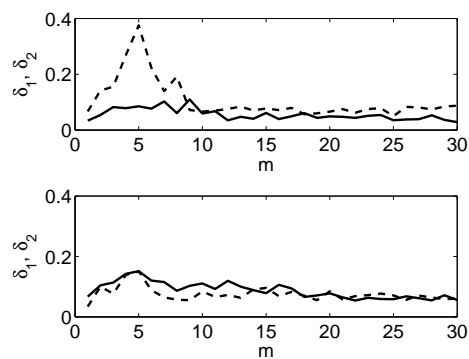,height=5.cm}
\end{center}
\caption{{\small $\delta_1$ (full line) and $\delta_2$ (dashed line) are plotted vs $m$,
averaged over  CHF patients (above), and averaged over sepsis patients (below).
\label{deltachf}}}
\end{figure}

Now we describe the application to time series of heart rate and blood pressure from
patients from an intensive care unit, contained in the MIMIC database \cite{physionet}.
In healthy subjects the  heart rate variability (HRV) and the systolic blood pressure
(SBP) are interdependent. Two mechanisms determine the feed-forward influence of HRV on
SBP, the Starling law and the diastolic decay; baroreflex regulation determines the
influence of SBP on HRV \cite{koepchen}. We consider signals from six patients affected
by congestive heart failure(CHF)/pulmonary edema (patients no. $212$, $213$, $214$,
$225$, $230$, and $245$) and six patients whose primary pathology was sepsis,  a
condition in which the body is fighting a severe infection and that can lead to shock, a
reaction caused by lack of blood flow in the body (patients no. $222$, $224$, $269$,
$291$, $410$, and $422$). HRV and SBP time series are extracted from  raw data and
re-sampled at $2 Hz$. We fix $n=20$ and vary $m$ from $1$ to $20$, taking
$\sigma=2.5/\sqrt{m}$. Denoting  $x$ the HRV time series, and $y$ the SBP time series,
figure \ref{figepsvsm} shows the behavior of $\epsilon_x$ and $\epsilon_y$ as a function
of $m$ for a typical subject. The optimal value of $m$ corresponds to the knee of the
curves, $m=5$: in terms of frequency this value corresponds to the respiratory band. In
figure \ref{deltachf} we depict $\delta_1=(\epsilon_x - \epsilon_{xy})/\epsilon_x$
(measuring the influence of SBP on HRV) and $\delta_2=(\epsilon_y -
\epsilon_{yx})/\epsilon_y$ (measuring the influence of HRV on SBP), as a function of $m$,
for CHF and sepsis patients. In the case of sepsis patients, the curves show a symmetric
HRV-SBP interdependence, whilst in CHF patients the $HRV\to SBP$ influence seems to be
dominant, as $\delta_2$ shows a peak at $m=5$. The probability that the twelve $\delta_2$
values, from all subjects and corresponding to $m=5$, were drawn from the same population
has been estimated by Wilcoxon test to be less than $10^{-2}$. The average directionality
index $D$ \cite{ancona}, at $m=5$, is equal to $0.82$ for CHF patients and to $-0.019$
for sepsis patients. These results show, on one side, that sepsis condition is not
characterized by  unbalanced HRV-SBP loop interaction. On the other hand, it is well
known that CHF patients show unbalanced HRV-SBP regulatory mechanism, the feed-forward
HRV$\to$SBP coupling being prevalent over baroreflex sensitivity \cite {pin}: our
findings show that this effect may be described in terms of Granger causality between HRV
and SBP time series.

\begin{figure}[ht!]
\begin{center}
\epsfig{file=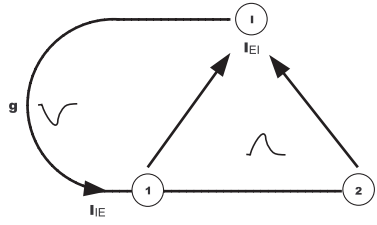,height=5.cm, angle=270}
\end{center}
\caption{{\small  Neural model architecture. \label{figmodel}}}
\end{figure}

\begin{figure}[ht!]
\begin{center}
\epsfig{file=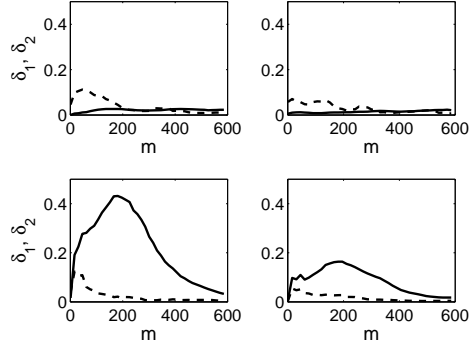,height=5.cm}
\end{center}
\caption{{\small $\delta_1$ (full line, $I\rightarrow E$) and
$\delta_2$ (dashed line, $E\rightarrow I$) are plotted vs $m$, for
$\tau_I$ = $\tau_E$ =$6 ms$ (left), and $\tau_I$ = $18$ ms, $\tau_E$
=$6 ms$ (right). Above, no feedback. Below, with feedback.
\label{deltan} }}
\end{figure}

As a second application, we consider the interactions in a model of
inhibitory and excitatory neurons, which has been studied from
different points of view, and with different aims. Much attention
have been dedicated to the mechanisms underlying modulation of
visual processing by means of attention \cite{reynolds}.
Synchronization originating from feedback can be one way to explain
this process of selective attention. In a previous study
\cite{Doiron}, it has been shown that inhibitory feedback is a way
for the neurons to distinguish between spatially correlated and
uncorrelated input. So it is important to study causal influences in
a closed feedback-feedforward loop, and to understand how the
inhibitory feedback, responsible of the selectivity of the
attention, modifies the causal relationships between the neurons in
the excitatory population. Here we consider the following question:
what does causality measure in the case of coupled firing neurons?
We show that it does not merely measure coupling constants, but the
combined influence of couplings and membrane time constants. We
consider a model of interacting neurons, two excitatory and one
inhibitory, as illustrated in figure \ref{figmodel}. This is a
simplified case of the models in \cite{Doiron}, \cite{danstan}  and
\cite{Kopell}. All neurons are Leaky Integrate-and-Fire (LIF)
neurons with a membrane potential $V$ and an input current $I$
satisfying:
\begin{equation}\label{lifmemb}
\dot{V} = -V+I,
\end{equation}
where time is measured in units of the membrane time constant and
the membrane resistance is set to one. We denote $\tau_E$ ($\tau_I$)
the membrane time constant of excitatory (inhibitory) neurons. Every
time the potential of a neuron reaches the threshold value $V_{th}$,
a spike is fired. This resets the potential to the value $V_{R}$; it
remains bound to this value for an absolute refractory period
$\tau_{R}$. Furthermore we normalize the reset value $V_{R}$ to zero
and the threshold value $V_{th}$ to one. The series of spikes
$S^E_j(t)$ ($j=1,2$) from excitatory neurons provide the input
current $I_I$ to the inhibitory LIF neuron, given by the convolution
of the sum of the spike trains of the excitatory neurons and a
standard $\alpha$ function, which mimics an excitatory post-synaptic
potential (EPSP). As a consequence of the excitatory input, the
inhibitory neuron fires action potentials. As in the case of the
excitatory current, the action potential $S^I(t)$ of the inhibitory
neuron provides the input currents $I_{E,j}$ to the excitatory
neurons by convolution of the spike train from the inhibitory neuron
and an inhibitory post-synaptic potential (IPSP), given by the same
$\alpha$ function used to represent the EPSP, but reversed in sign.
This inhibitory feedback is characterized by a gain $g$, and it's
delivered after a delay time $\tau_{D}$. Summarizing, for each
excitatory neuron $j=1,2$ the total input current $I_{E,j}$ is given
by
\begin{equation}
\label{inputexc} I_{E,j}(t)=\mu + \eta_{j}(t) - g \int_0^{\infty}d\tau s^I(t)(t-\tau -
\tau_D )\alpha^{2}\tau e^{-\alpha \tau},
\end{equation}
while the input current for the inhibitory neuron is
\begin{equation}
\label{inputinh} I_I(t)=\mu + \int_{0}^{\infty}d\tau
\left(S^E_1(t-\tau)+S^E_2(t-\tau)\right)\alpha^{2}\tau e^{-\alpha \tau},
\end{equation}
where $\mu$ is a constant base current and $\eta_{j}(t)$ represents internal Gaussian
white noise with intensity $K$. We use $\mu$ = $0.5$, $K$ = $0.08$, $\alpha$ = $18$ ms
and $\tau_{D}$ = $18$ ms. $\tau_{R}$ is set to $3$ ms, while $\tau_E =\tau_I =6$ ms. Note
that the base current is below threshold, therefore the inhibitory neuron only fires in
response to excitatory spike input. Moreover, using these parameters, the feedback model
is a very stable one, resulting in firing rates from all neurons being almost independent
on the values of membrane time constants and inhibitory feedback gain $g$ used in this
study. The firing rates of all the three neurons are always between $40$ and $60$ Hz.
The model equations have been numerically solved by Eulero integration using a time step
0.1 ms. The causal relationships between the time series of  membrane potentials from
neurons are then evaluated using the proposed RBF approach with $n=20$ and
$\sigma=2.5/\sqrt{m}$, $m\in [1,600]$. In absence of inhibitory feedback, i.e. $g=0$,
there is a dominant causal influence of the excitatory input on the inhibitory one; on
the other hand, when the feedback is turned on (g = $1.2$), the causal influence is
reversed, see figure \ref{deltan}. It is worth stressing, however, that in this context
the measure of causality is not simply related to coupling $g$. Indeed, if we consider
the same situation, but with a longer membrane time constant of the inhibitory neuron
$\tau_I$, the influence of the latter on the excitatory ones is notably reduced (see
figure \ref{deltan}). This can be explained in the following way. When the time constant
is short, the inhibitory neuron behaves as a coincidence detector and inhibitory feedback
has the effect of synchronizing the excitatory neurons \cite{Kopell}, which thus tend to
fire at the same time. For longer membrane time constant, the inhibitory neuron acts as
an integrator and does not discriminate between two temporally close spikes; the
inhibitory neuron then reacts to a smaller number of input spikes, resulting in a
decreased causal influence on the two excitatory neurons.

\begin{figure}[ht!]
\begin{center}
\epsfig{file=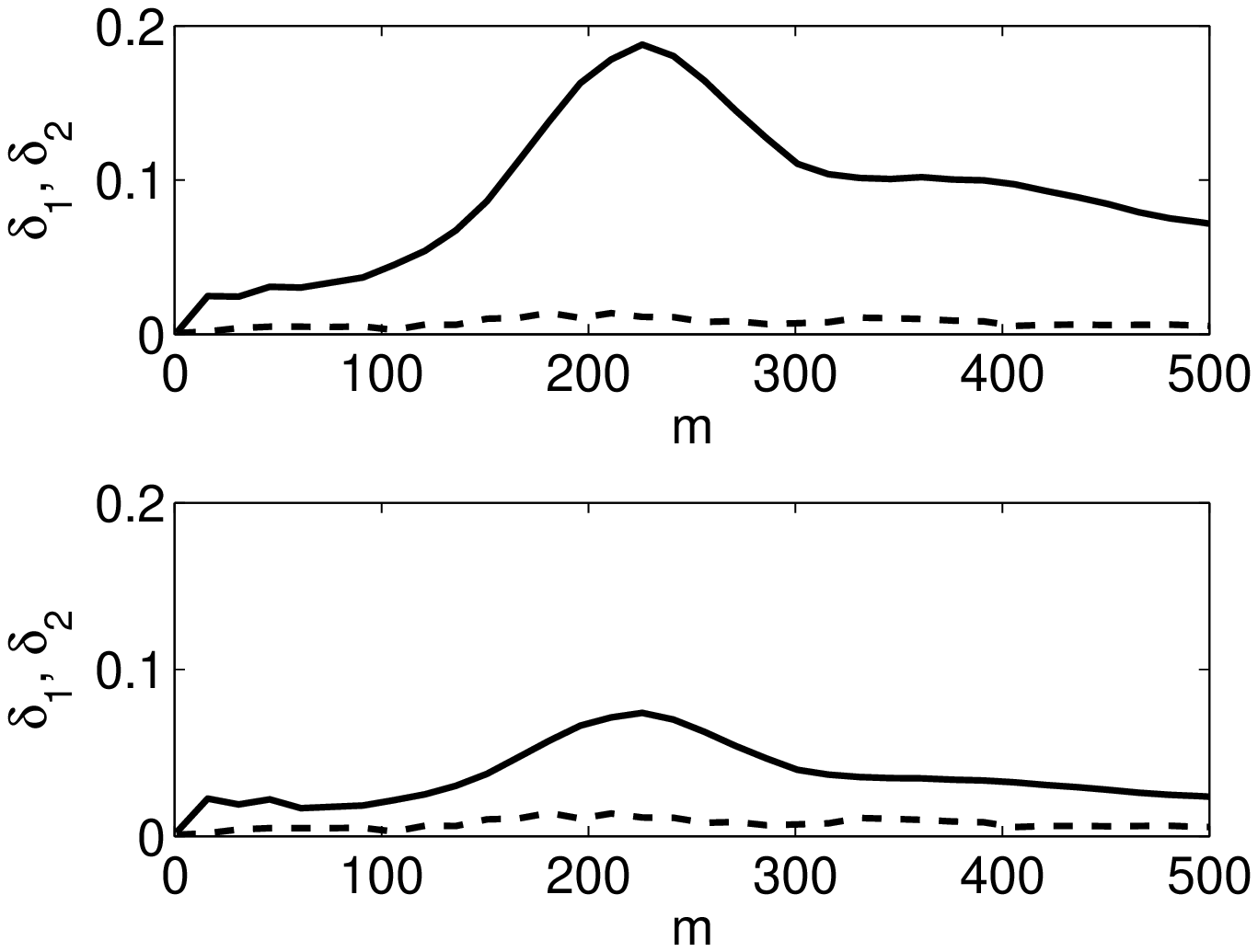,height=5.cm}
\end{center}

\caption{{\small $\delta_1$ (full line, influence of the excitatory neuron without
feedback on the one with feedback) and $\delta_2$ (dashed line, influence of the
excitatory neuron with feedback on the one without feedback) are plotted vs $m$, for
$g=1.2$. $\tau_I$ = $\tau_E$ =$6 ms$ (top); $\tau_I$ = $18$ ms, $\tau_E$ =$6 ms$
(bottom). \label{delta2neur}}}
\end{figure}

Another interesting situation corresponds to only one, out of the two, excitatory neuron
receiving inhibitory feedback: we then investigate  the causal relationships between the
two excitatory neurons, as mediated by the inhibitory one. In absence of inhibitory
feedback there is no causality involved. As the feedback is switched on, as it is clear
in figure \ref{delta2neur}, a significant causal influence, of the neuron which does not
receive the feedback on the other, is observed. Even in this case, the causal influence
depends  also on the membrane time constant of the inhibitory neuron, which may act as a
coincidence detector or as an integrator, thus considering two successive spikes from the
two excitatory neurons as two separates spikes, or as one, respectively. This mechanism
of selective feedback can be used to explain the phenomenon of stimulus recognition with
biased attention in the visual cortex \cite{reynolds}.

We have generalized the RBF approach to Granger causality so that any function of the
input variables can be approximated. The two applications here considered show the
usefulness of the proposed approach and, more generally, of the notion of causality.

\vskip 0.4 cm\par\noindent{\bf Acknoledgements.} Valuable discussions with Nicola Ancona
and Stan Gielen are warmly acknowledged.


\begin{thebibliography}{99}
\bibitem{granger} C.W.J. Granger, Econometrica {\bf 37}, 424 (1969).
\bibitem{ting} J.J. Ting, Physica A {\bf 324}, 285 (2003).
\bibitem{tass} P. Tass et al., Phys. Rev. Lett. {\bf 81}, 3291 (1998); M. Le van Quyen et
al., Brain Res. {\bf 792}, 24 (1998).
\bibitem{arnold} S. J. Schiff et al., Phys. Rev. E {\bf 56}, 6708 (1996); M.
Wiesenfeldt, U. Parlitz, W. Lauterborn, Int. Jour. of Bifurcation
and Chaos {\bf 11}, 2217 (2001).
\bibitem{sowa} R. Sowa et al., Phys. Rev. E {\bf 71}, 61926 (2005).
\bibitem{blinowska} K.J. Blinowska, R. Kus and M. Kaminski, Phys. Rev. E {\bf 70},
50902(R) (2004).
\bibitem{rosemblum} M. G. Rosenblum et al., Phys. Rev. E {\bf 64}, 45202R (2001);
M. Palus, A. Stefanovska, Phys. Rev. E {\bf
67}, 55201R (2003).
\bibitem{pereda} E. Pereda, R. Quian Quiroga, J. Bhattacharya, Progress in Neurobiology {\bf 77} 1
(2005).
\bibitem{cardioresp} M. G. Rosenblum et al., Phys. Rev. E {\bf 65},
41909 (2002).
\bibitem{chen} Y. Chen et al., Phys. Lett. A {\bf 324}, 26 (2004).
\bibitem{ancona} N. Ancona, D. Marinazzo and S. Stramaglia, Phys. Rev. E {\bf 70}, 56221
(2004).
\bibitem{verdes} P.F. Verdes, Phys. Rev. E {\bf 72}, 26222 (2005).
\bibitem{nc} N. Ancona and S. Stramaglia, cond-mat/0502511 (in press in Neural
Computation).
\bibitem{nota1} Usually both times series are normalized in the preprocessing stage,
i.e. they are linearly transformed to have zero mean and unit variance.
\bibitem{fcm} J. C. Bezdek, {\em Pattern Recognition with Fuzzy Objective Function Algorithms\/} (Plenum Press, New York, 1981).
\bibitem{physionet} http://www.physionet.org/
\bibitem{koepchen}{\it Mechanisms of blood pressure waves}, K. Miyakawa, C. Polosa, H.P. Koepchen
(eds.). Springer, Berlin Heidelberg New York (1984).
\bibitem{pin} G.D. Pinna et al.,
J. Am. Coll. Cardiol. {\bf 46}, 1314 (2005).
\bibitem{reynolds} J.R. Reynolds et al., J. Neurosci. {\bf 19},
1736 (1999).
\bibitem{Doiron} B. Doiron et al., Phys. Rev. Lett. {\bf 93(4)},
048101 (2004).
\bibitem{danstan} D. Marinazzo, H.J. Kappen, S. Gielen, {\it Input-driven oscillations in
networks with excitatory and inhibitory neurons with dynamic synapses}, submitted.
\bibitem{Kopell} C. B\"{o}rgers  and N. Kopell,  Neural Computation {\bf 15},
509 (2003).

\end{thebibliography}
\end{document}